\documentclass[twocolumn,showpacs,%
  nofootinbib,aps,superscriptaddress,%
  prd,notitlepage,showkeys,10pt]{revtex4-1}

\usepackage{amssymb}
\usepackage{amsmath}
\usepackage{graphicx}
\usepackage{dcolumn}
\usepackage{hyperref}

\begin{document}

\title{Quantum Erasure Cryptography}
\author{Hatim Salih}
\email[]{salih.hatim@gmail.com}
\affiliation{Qubet Research, London NW6 1RE, UK}

\date{\today}

\begin{abstract}


The phenomenon of quantum erasure has long intrigued physicists, but has surprisingly found limited practical application. Here, we propose an erasure-based protocol for quantum key distribution (QKD) that promises inherent security against detector attacks.

\end{abstract}

\maketitle

\section{Introduction}

%
%
Ideas at the root of quantum erasure were already at play in the famous Bohr-Einstein dialogues in the 1920's. In one such debate, Einstein envisaged a scenario where one could seemingly observe interference fringes in a double-slit experiment, as well as learn which-path information for individual photons \cite{Bohr}. Bohr countered, arguing that Heisenberg's uncertainty principle would prevent one from doing exactly that. But as it later turned out, the deeper explanation was in fact not uncertainty but rather entanglement---a concept Einstein himself helped put on the map in the EPR paper \cite{EPR}.

In 1982 Scully and Druhl \cite{Scully and Druhl} proposed a variant of the double-slit experiment that sent shock-waves through the physics community. The idea was that which-path information can be inferred without disturbing the trajectory of individual particles, circumventing the uncertainty argument. A simpler, subsequent proposal used excited atoms \cite{Scully et al.}. Scully and colleagues showed that interference would still be lost due to entanglement providing a which-path ``tag''. In the latter proposal, the presence of an emitted photon in one of two microwave cavities placed directly behind the two slits (with the cavities stretching long enough for the atoms to drop from their exited state) would indicate which slit an atom went through. However, erasing this which-path information, by removing a wall between the two cavities, would restore the interference fringes: quantum erasure.

While the original quantum erasure proposals have not been implemented, other more practical ones have. For instance, Walborn et al. \cite{Walborn} directed photons, each from a polarisation-entangled pair, towards a double-slit. Each slit had a polarisation rotator that imprinted a which-path tag. Measuring the polarisation of the other photon of the entangled pair in the right way erases which-path information. Here, interference fringes can even be recovered long after the photons passing through the double-slit have been detected.

Despite its fundamental significance, quantum erasure has for one reason or another struggled for practical application. One notable exception is Zhao et al. recently employing a frequency eraser to entangle, for the first time, different-color photons \cite{Zhao}.

Quantum key distribution (QKD) on the other hand enables two remote parties to share a random string of zeros and ones. Given such a sting, provably secure communication can be established \cite{Shannon}. Much of quantum cryptography thus reduces to QKD. In 1984, Bennett and Brassard proposed the first QKD protocol, the BB84 \cite{BB84}, which has since been shown to be unconditionally secure against an eavesdropper with unlimited resources \cite{Shor}. However, imperfect devices, especially imperfect detectors, allow powerful so called side-channel attacks that can compromise the security of QKD. Lydersen et al. \cite{Lydersen}, for instance, showed how the secret key in two commercially available QKD systems can be fully obtained using their detector-blinding attack, exploiting a common detector imperfection.  

\section{Methods}

\begin{figure}
\centering
\includegraphics[width=0.5\textwidth]{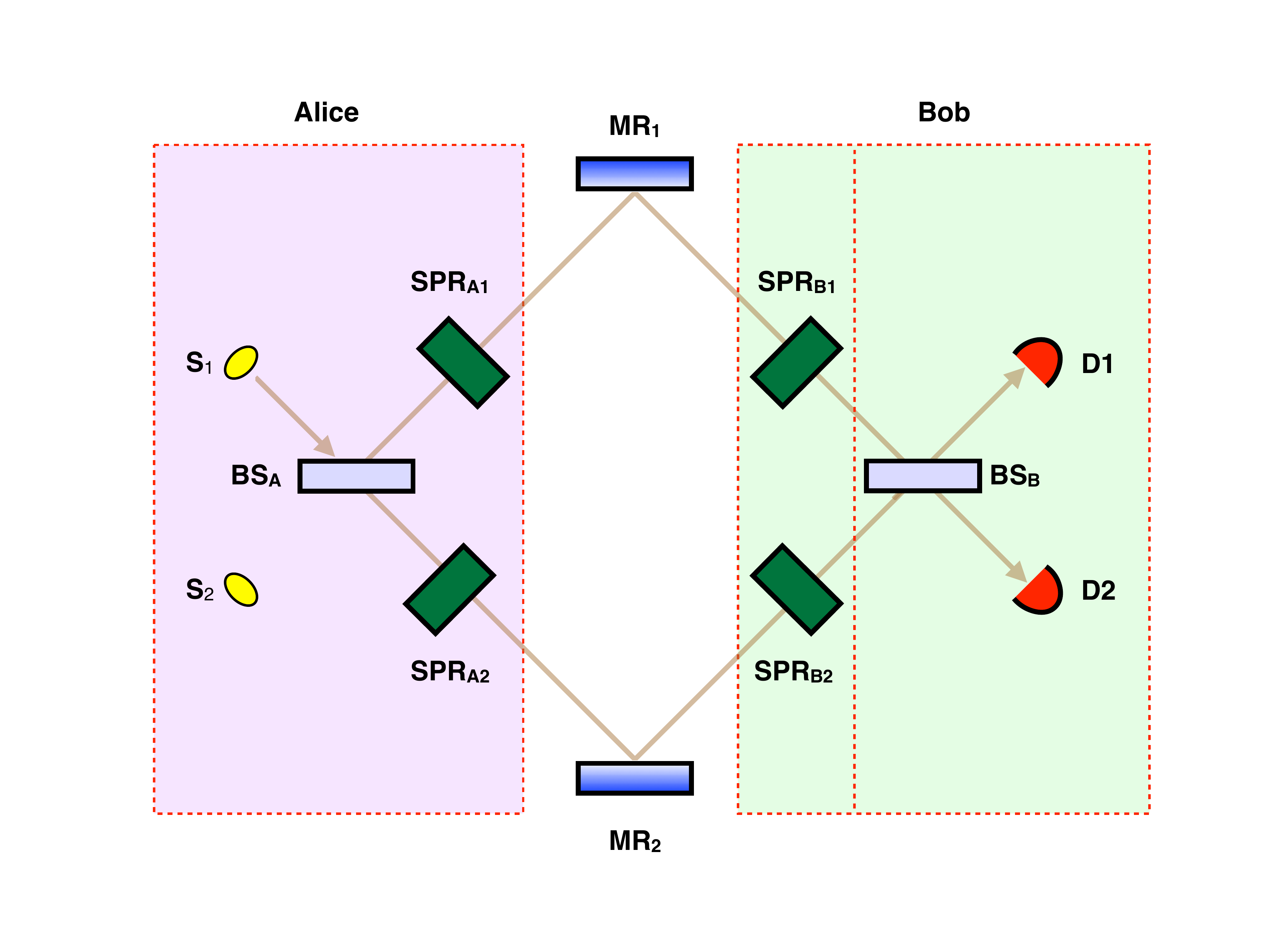}
\caption{\label{fig: Figure}Quantum erasure cryptography. After sending in her photon as shown, Alice encodes bit value `0' by doing nothing, and encodes bit value `1' by applying switchable polarisation rotators $SPR_{A1}$ and $SPR_{A2}$. Bob encodes bit value `1' by doing nothing, and encodes bit value `0' by applying switchable polarisation rotators $SPR_{B1}$ and $SPR_{B2}$. Which-path information destroys interference, while the erasure of which-path information restores interference. As explained in the text, for Alice sending in her photon from the top left, $D_1$ clicking uniquely corresponds to Alice and Bob agreeing in their bit choices. $S$s are single-photon sources. $MR$s are mirrors. $BS$s are 50-50 beam-splitters. Practically, the upper and lower arms of this interferometer would be implemented using two optical cables, in which case there would be no need for $MR$s.}
\end{figure}

We now describe our erasure-based protocol for QKD that promises security against detector attacks. The goal here for ``Alice'' and ``Bob'', the customary communicating parties in such tasks, is to securely share a random string of zeros and ones. We start by explaining a simplified two-state version of the protocol before proceeding to give the complete four-state protocol. After sending her photon from the top left, Alice encodes the bit value `0' by doing nothing, and encodes the bit value `1' by turning on switchable polarisation rotators $SPR_{A1}$ and $SPR_{A2}$, applying rotations $R(-\pi/2)$ and $R(\pi/2)$ to photon polarisation in upper and lower paths respectively, as shown in Fig. \ref{fig: Figure}. Bob on the other hand encodes the bit value `1' by doing nothing, and encodes the bit value `0' by turning on switchable polarisation rotators $SPR_{B1}$ and $SPR_{B2}$, applying rotations $R(\pi/2)$ and $R(-\pi/2)$ to photon polarisation in the upper and lower paths respectively.

More precisely, Alice starts by sending in a photon, using single-photon source $S_1$, in the state $\left| \text{0} \right\rangle \left| +45 \right\rangle$, where $\left| \text{0} \right\rangle$ corresponds to the photon being in the upper path, and $\left| +45 \right\rangle$ corresponds to $45^{\circ}$ polarisation. After passing through Alice's 50-50 beamsplitter, the state becomes $\frac{1}{\sqrt{2}}(\left| \text{0} \right\rangle + \left| \text{1} \right\rangle) \left| +45 \right\rangle$. If Alice chooses the bit value `0', not applying her rotations, the state remains unchanged. If Alice chooses the bit value `1', applying her rotations, the state becomes $\frac{1}{\sqrt{2}}\left| \text{0} \right\rangle \left| H \right\rangle + \frac{1}{\sqrt{2}}\left| \text{1} \right\rangle \left| V \right\rangle$. See Fig. 1. Crucially, the two states $\frac{1}{\sqrt{2}}(\left| \text{0} \right\rangle + \left| \text{1} \right\rangle) \left| +45 \right\rangle$ and $\frac{1}{\sqrt{2}}\left| \text{0} \right\rangle \left| H \right\rangle + \frac{1}{\sqrt{2}}\left| \text{1} \right\rangle \left| V \right\rangle$ are nonorthogonal and therefore cannot be reliably distinguished by Eve \cite{Nielsen and Chuang}.

For the case of Alice choosing the bit value `0', sending $\frac{1}{\sqrt{2}}(\left| \text{0} \right\rangle + \left| \text{1} \right\rangle) \left| +45 \right\rangle$ into the channel, there is no which-path information. If Bob does nothing, choosing the bit value `1', $D_2$ clicks with certainty because of destructive interference at $D_1$. But if Bob applies his rotations, choosing the bit value `0', then which-path information is introduced in the form of entanglement in $\frac{1}{\sqrt{2}}\left| \text{0} \right\rangle \left| H \right\rangle + \frac{1}{\sqrt{2}}\left| \text{1} \right\rangle \left| V \right\rangle$. No interference takes place: $D_1$ and $D_2$ are equally likely to click.

For the case of Alice choosing the bit value `1', sending $\frac{1}{\sqrt{2}}\left| \text{0} \right\rangle \left| H \right\rangle + \frac{1}{\sqrt{2}}\left| \text{1} \right\rangle \left| V \right\rangle$ into the channel, there is which-path information. If Bob applies his rotations, choosing the bit value `0', then which-path information is erased, resulting in the state $\frac{1}{\sqrt{2}}(\left| \text{0} \right\rangle + \left| \text{1} \right\rangle)\left| +45 \right\rangle$. $D_2$ clicks with certainty because of destructive interference at $D_1$. But if Bob does nothing, choosing the bit value `1', we get no interference: $D_1$ and $D_2$ are equally likely to click. Importantly, whenever $D_1$ clicks, Alice and Bob have agreed in their bit choices. Alice publicly instructs Bob to keep the corresponding bits, which form our sifted key. Table \ref{tab:detector action} shows Alice and Bob's random bit choices and corresponding detector action.

This simplified two-state version is vulnerable to a simple Eve attack. She can make the exact measurement as Bob using two polarisation rotators, a 50-50 beamsplitter, and two detectors. For Eve not applying her rotations, whenever her equivalent of Bob's $D_1$ clicks, she knows Alice has sent her entangled state; Eve sends the entangled state. If Eve's other detector clicks, she sends nothing. The case of Eve applying her rotations is analogous. This way Eve can obtain the full key.

We are ready to give our quantum erasure cryptography protocol, using four states. Alice now sends her photon either from the top using $S_1$, or from the bottom using $S_2$, with equal probability. After Bob makes his measurement, Alice announces publicly which photons were sent from the top and which from the bottom. For photons sent from the top, as before, bits corresponding to $D_1$ clicking are kept while the rest are thrown away. For photons sent from the bottom, bits corresponding to $D_2$ clicking are kept while the rest are thrown away. More precisely, the protocol proceeds as follows:

\begin{enumerate}

\item With probability $1/4$ Alice sends Bob one of four possible states: $\frac{1}{\sqrt{2}}(\left| \text{0} \right\rangle + \left| \text{1} \right\rangle) \left| +45 \right\rangle$ or $\frac{1}{\sqrt{2}}(\left| \text{1} \right\rangle - \left| \text{0} \right\rangle) \left| +45 \right\rangle$, which are both unentangle, or $\frac{1}{\sqrt{2}}\left| \text{0} \right\rangle \left| H \right\rangle + \frac{1}{\sqrt{2}}\left| \text{1} \right\rangle \left| V \right\rangle$, or $\frac{1}{\sqrt{2}}\left| \text{1} \right\rangle \left| V \right\rangle - \frac{1}{\sqrt{2}}\left| \text{0} \right\rangle \left| H \right\rangle$, which are both entangled. She encodes bit value `0' by sending Bob one of the unentangled states, and bit value `1' by sending one of the entangled states.
\item Bob encodes bit value `0' by making the measurement corresponding to applying his polarisation rotators, and encodes bit value `1' by making the measurement corresponding to not applying his polarisation rotators.
\item Alice announces whether she initially sent her photon from the the top using $S_1$ or from the bottom using $S_2$.
\item Bob announces which of his two detectors clicked for each photon.
\item For photons sent from the top using $S_1$, bits corresponding to $D_1$ clicking are kept. For photons sent from the bottom using $S_2$, bits corresponding to $D_2$ clicking are kept instead. Those bits that are kept form the sifted key. The rest are thrown away.
\item Alice randomly chooses a sample from the sifted key. Alice and Bob publicly announce corresponding bits. If error rate exceeds some threshold they then abort protocol and start over.
\item Alice and Bob perform error correction and privacy amplification to obtain the final secure key.

\end{enumerate}
 
One of the reviewers of the present paper brought to our attention a one-qubit protocol, reference \cite{Xiongfeng}, that shares an important feature with our two-qubit protocol; the secure key is extracted from measurement choices rather than measurement outcomes. Whereas for both protocols three out of four photons are on average thrown away, in our two-qubit protocol two out of the three photons that are to be thrown away can be used to check for an attack by Eve as shown in the next section.
 
\section{Results and discussion}

We discuss two attacks by Eve. First an intercept-resend attack, then a powerful detector-blinding attack. Consider Eve employing an intercept-resend strategy where, just like Bob, she brings the two paths together, randomly choosing to either apply identical rotations to Bob's or not apply any before her beam-splitter and two detectors which are also identical to Bob's. Take the case of Eve not applying her rotations and her lower detector clicking. The detector could have equally been triggered by an unentangled state or by an entangled one. But one of the two unentangled states, which allow interference to take place, can be ruled out. Therefore by assuming that Alice has sent the other entangled state, in this case the one corresponding to Alice sending in her photon from the top and not applying her rotations, Eve is correct with probability $1/2$. She sends this state to Bob. The case of Eve not applying her rotations and her upper detector clicking is analogous; she sends to Bob Alice's other unentangled state, that is the one corresponding to Alice sending in her photon from the bottom and not applying her rotations. For the case of Eve applying her rotations, if her lower detector clicks she sends Alice's entangled state corresponding to Alice sending in her photon from the top and applying her rotations. And if Eve's upper detector clicks she sends Alice's entangled state corresponding to Alice sending in her photon from the bottom and applying her rotations. 

Let's now work out the probability of Bob's detector $D_1$ incorrectly clicking for the case of Alice sending in her photon from the top. Eve sends the wrong state with probability $1/2$. Based on Alice and Bob's bit choices, there is a $1/2$ chance that the state incident on Bob's beam-splitter should be the unentangled state, corresponding to Alice and Bob not agreeing on their bit choices, leading to destructive interference at Bob's $D_1$. The probability of Eve incorrectly triggering $D_1$ is therefore $1/2 \times 1/2 \times 1/2=1/8$. The probability of Eve correctly triggering $D_1$ in this case is $1/4$, with Alice and Bob agreeing on their bit choices. ($1/8$ due to Eve sending correct state plus $1/8$ due to Eve sending wrong state.) By symmetry, The probability of Eve incorrectly triggering $D_2$ for the case of Alice sending in her photon from the bottom is also $1/2 \times 1/2 \times 1/2=1/8$. The probability of Eve correctly triggering $D_2$ in this case is $1/4$, with Alice and Bob agreeing on their bit choices. ($1/8$ due to Eve sending correct state plus $1/8$ due to Eve sending wrong state.) Given this attack, the error rate in the sifted key, called the quantum bit error rate or QBER, is $1/8 \div 3/8 = 1/3$.

We now show that our protocol is secure against this intercept-resend attack. Starting with the sifted key, the classical algorithms of error correction and privacy amplification can be used to generate a secure key---as long as Bob has more information than Eve \cite{Gisin et al.}. More precisely, Bob's mutual Shannon information $I(\alpha,\beta)$ has to be greater than Eve's mutual Shannon information $I(\alpha,\epsilon)$ which is given by,

\begin{equation}
I(\alpha,\epsilon) = 1+\sum_{r=0,1,0',1'} P(r)\sum_{i=0,1}\frac{P(r|i)P(i)}{P(r)}\log_{2}\frac{P(r|i)P(i)}{P(r)}
\end{equation} 

where $P(i)$ is the probability of Alice sending bit $i$ to Bob, with $i$ being either 0 or 1. $P(r)$ is the probability of Eve getting measurement outcome $r$, with $r$ being either 0 or 1 for Eve not applying her rotations, or $0'$ or $1'$ for Eve applying her rotations, where 0, $0'$ correspond to bit value `0', and 1, $1'$ correspond to bit value `1'. $P(r|i)$ is the probability of Eve getting $r$ given Alice's bit $i$. With Eve's strategy, $P(r=0)$ is $3/8$, $P(r=1)$ is $1/8$, $P(r=0')$ is $1/8$, $P(r=1')$ is $3/8$, $P(i=0)$ is $1/2$, $P(i=1)$ is $1/2$, $P(r=0|i=0)$ is 1/2, $P(r=1|i=0)$ is 0, $P(r=0'|i=0)$ is $1/4$, $P(r=1'|i=0)$ is $1/4$, $P(r=0|i=1)$ is 1/4, $P(r=1|i=1)$ is $1/4$, $P(r=0'|i=1)$ is 0, $P(r=1'|i=1)$ is 1/2. Eve's information gain $I(\alpha,\epsilon)$ is therefore 0.311. Alice and Bob randomly choose a sample of the sifted key, which would be thrown away. If for this sample the estimated QBER $<$ $33.3\%$, and the estimated $I(\alpha,\beta)>0.311$, they proceed with the classical algorithms of error correction and privacy amplification to generate their secure key, otherwise they stop the protocol.


While for Bennett and Brassard's BB84 QKD protocol \cite{BB84}, for comparison, only one out of two photons is thrown away in order to get the sifted key, in our protocol three out of four photons are thrown away. For this price, however, Eve's QBER in our erasure-based protocol is higher than that for BB84, and Eve's information gain is lower. For a typical intercept-resend strategy by Eve on BB84, Eve's QBER is $25\%$ for an information gain $I(\alpha,\epsilon)$ of $1/2$, compared to a QBER of $33.3\%$ for an information gain $I(\alpha,\epsilon)$ of 0.311 in our protocol.

We have mentioned that Alice and Bob can make use of two out of the three photons that are to be thrown away on average, to check for an attack by Eve. When Alice and Bob make opposite bit choices, which means interference should take place, Bob measures polarisation in the $+45$, $-45$ basis. He should always measure $+45$. (Bob can make this measurement by replacing each of his detectors by a $45^{\circ}$ polarisation rotator, which leads to a polarising beam-splitter, which in turn leads to two detectors.) For Eve's intercept-resend attack above, Eve sends the wrong state half the time on average. This means Bob would measure $-45$ polarisation with $25\%$ probability, alerting him to Eve's attack.    

Let us now look at Eve's detector-blinding attack inspired by reference \cite{Lydersen}. Eve performs the same measurement as in the intercept-resend attack discussed above. Her information gain $I(\alpha,\epsilon)$ therefore remains the same, 0.31. By means of high intensity light, Eve blinds Bob's detectors, causing them to only click for pulses of light with intensity above a given threshold. With probability $1/2$ Eve correctly guesses Alice's state. Say Eve guesses that Alice sent an unentangled state, that is bit value `0'. She sends a strong enough pulse made up entirely of only one of Alice's two entangled states chosen such that if Bob's bit is also `0', which corresponds to Bob applying his rotations, the whole pulse will end up at $D_1$($D_2$) for the case of Alice sending her photon from top(bottom). If on the other hand Bob chooses bit value `1', which corresponds to him not applying his rotations, roughly half of Eve's signal would go to one detector while the other half would go to the other detector. The intensity of Eve's pulse is chosen such that half its intensity is below the threshold for triggering either detector. No detector clicks. But Eve sends the wrong state to Bob roughly half the time. And given that Alice and Bob choose different bits half the time, the probability of Eve incorrectly triggering $D_1$($D_2$) for the case of Alice sending in her photon from the top(bottom) is $1/2 \times 1/2 \times 1/2=1/8$. The probability of Eve correctly triggering $D_1$($D_1$) is $1/2 \times 1/2=1/4$. The QBER is therefore $1/8 \div (1/4+1/8) = 1/3$. As before, for their sample of the sifted key, if estimated QBER $<$ $33.3\%$, and estimated $I(\alpha,\beta)>0.311$, Alice and Bob proceed with the classical algorithms of error correction and privacy amplification to generate their secure key, otherwise they stop the protocol. By comparison, against the BB84, the detector-blinding attack enables Eve to obtain the full secret key without introducing any errors. 

What about attacks whose aim is to learn which of Bob's detectors clicked? Because for each photon Bob publicly announces which detector clicked, if any, such attacks by Eve are not relevant. In fact, no bit-value information is encoded in the photon incident on Bob's beam-splitter. Bob's beam-splitter and two detectors, it seems, might as well be handed to Eve---as long as the module containing his two polarisation rotators is kept secure. A general proof of the security of our protocol is planned for a separate paper. 

In summary, we have proposed an erasure-based protocol for quantum key distribution that promises inherent security against side-channel detector attacks, discussing its security against Eve's powerful detector-blinding attack.



\section*{Acknowledgments}
The author thanks M. Al-Amri, Sam L. Braunstein and Zeng-Hong Li for valuable discussions in 2013, and more recently Roger Colbeck. Qubet Research is a start-up in quantum information.

\clearpage



\begin{table}
    \begin{tabular}{|c|c|c|}
	\hline
	Alice's bit choice & Bob's bit choice & Detector clicking\\
	\hline
    0 & 0 & $D_{1}$ or $D_{2}$\\
	\hline
    0 & 1 & $D_{2}$\\
	\hline
    1 & 0 & $D_{2}$\\
	\hline
    1 & 1 & $D_{1}$ or $D_{2}$\\
	\hline
    \end{tabular}
    \caption{\label{tab:detector action}Detector action for different bit choices by Alice and Bob for the case of Alice sending in her photon from the top  as shown in Fig. \ref{fig: Figure}. $D_{1}$ clicking uniquely corresponds to Alice and Bob making the same bit choices.}
\end{table}

\end{document}